# Making Liquid Oxygen

M. M. J. French (1,2) and Michael Hibbert (2)

(1) INSPIRE, Imperial College, London
(2) The Ellen Wilkinson School for Girls, Acton, London

E-mail: mail@matthewfrench.net

In this article I explain in detail a method for making small amounts of liquid oxygen in the classroom if there is no access to a cylinder of compressed oxygen gas. I also discuss two methods for identifying the fact that it is liquid oxygen as opposed to liquid nitrogen.

Oxygen is one of the component gases of air at room temperature making up around 20% of the atmosphere. But can we liquify oxygen? As oxygen boils at 90 K /-183 °C and liquid nitrogen boils at 77 K /-196 °C we can use liquid nitrogen to liquify oxygen. The usual way to do this is to pass oxygen gas from a compressed gas cylinder through a coil of hollow copper pipe which is submerged in liquid nitrogen. The copper coil is a good conductor of heat and has a large surface area. Liquid oxygen is then usually collected in a thermos flask. However, the lack of a cylinder of compressed gas does not prevent liquid oxygen being made.

Liquid oxygen can be identified in two ways. Firstly pure liquid oxygen is light blue in colour, and secondly, liquid oxygen is paramagnetic. This means that it is affected by a magnetic field and is attracted to a sufficiently strong magnet.

Figure 1 shows the apparatus used to liquify oxygen without a compressed gas cylinder. A chemical reaction between manganese (IV) oxide (also known as manganese dioxide or $MnO_2$) and hydrogen peroxide ($H_2O_2$) produces oxygen gas in the conical flask (labeled A). Tracing the orange rubber hose to the left there is a boiling tube trap (labeled B) to prevent any hydrogen peroxide accidentally being taken into the coil. Again tracing the orange rubber hose to the left there is a copper coil submerged in a liquid nitrogen bath (labeled C). The top of the bath is also covered with foam/polystyrene for insulation. The outlet from the coil is connected to a second boiling tube trap where the liquid oxygen collects (labeled D). This is in a liquid nitrogen bath made using a Pyrex beaker which prevents the liquid oxygen boiling off. The rubber hose on the left is connected to a water tap vacuum pump to draw oxygen through the apparatus.

The main piece of specialist equipment used is the copper coil. It is made from 6mm diameter copper pipe bent 6 times around a suitable cylindrical object (eg a scaffold pole) around 7.5cm in diameter. If a compressed gas cylinder is used this coil is submerged in a bucket of liquid nitrogen from the top. By contrast, in this method the coil needs to be operated 'up side down' as there is not the pressure difference present with a compressed gas cylinder to force the liquid up and out of the pipe. Thus, the coil is placed into an expanded polystyrene bowl - basically a piece of packaging, as shown in figure 2. The small holes in the polystyrene which the coil was inserted though were sealed using 2 part epoxy. Check the epoxy active ingredients list for bisphenol-A epichlorohydrin as this seems to withstand the cold temperatures associated with liquid nitrogen very well.

After leaving the apparatus running for around 10-15 minutes a small amount of liquid oxygen had collected in the bottom of the boiling tube. This can be confirmed by the liquid being dragged up the side of the boiling tube by a neodymium magnet as well as a hint of light blue colour, see figure 3. The clamp stands can be arranged so that the nitrogen bath can be quickly removed and the liquid oxygen viewed before the boiling tube frosts up or it boils off. However, despite being attracted to the magnet, the blue colour is not as obvious as if a cylinder of compressed gas is used. I suspect this is due to some nitrogen contamination as the apparatus is not air tight.

Safety considerations need to be taken very seriously in performing this demonstration. Firstly, it should only be performed as a demonstration by a competent teacher who has gained confidence with the equipment by prior practice. Safety with liquid nitrogen needs to be considered: safety goggles should be worn at all times and sturdy leather/suede gloves should be worn when handling objects which have come into contact with liquid nitrogen and thus may be very cold. Liquid nitrogen should only be used or stored in a large/well ventilated room. Further safety advice on handling liquid nitrogen is available from CLEAPSS. Liquid oxygen is potentially more dangerous than liquid nitrogen. In addition to its very low temperature (so again goggles and gloves are required), liquid oxygen can cause materials which normally burn in oxygen to burn more fiercely. Therefore, care should be taken to ensure nothing which may burn (i.e. react with oxygen) can come into contact with the liquid oxygen. In general CLEAPSS recommends against handling liquid oxygen. However, provided all glassware is thoroughly clean (to remove any trace of materials which could react with liquid oxygen) and pupils are kept as far back as possible, a sufficiently thoughtful risk assessment and detailed series of safety precautions should allow this demonstration to be performed safely since only a few cubic centimeters of liquid oxygen are produced and quickly evaporate.

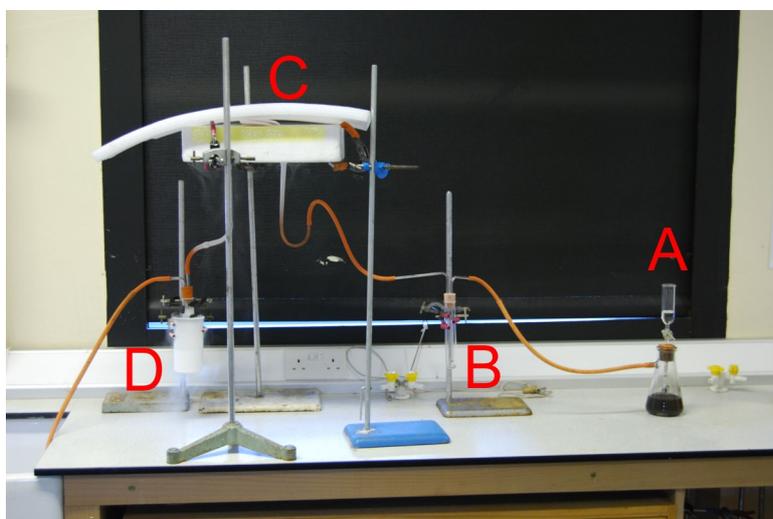

Figure 1: Apparatus for making liquid oxygen. Points A, B, C and D are referred to in the text.

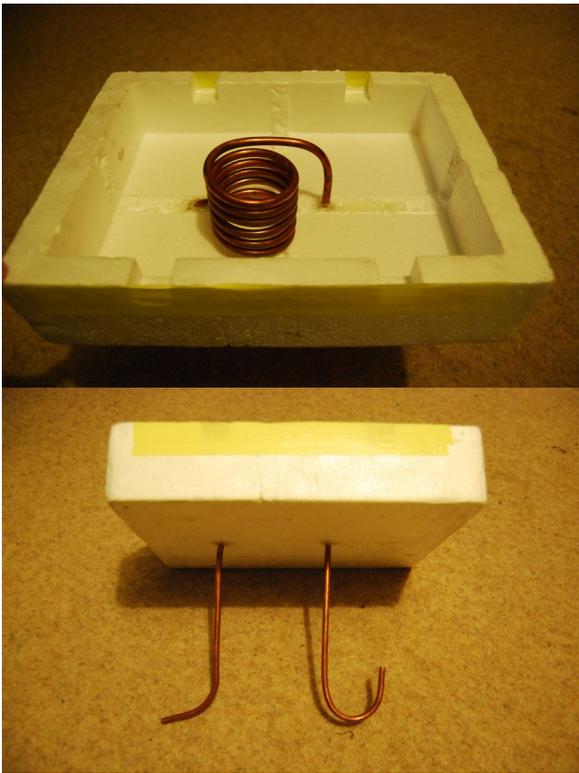

Figure 2: Copper coil and expanded polystyrene liquid nitrogen bath.

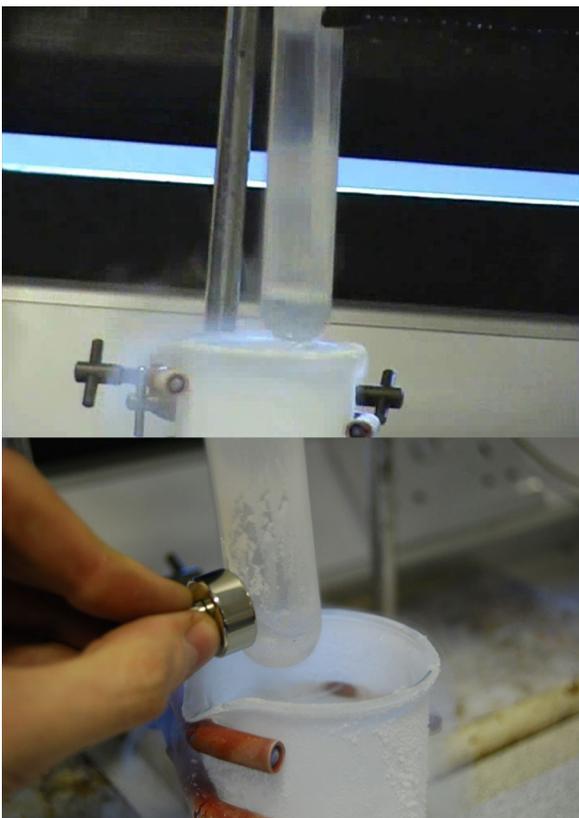

Figure 3: The top panel shows the slight blue colour of liquid oxygen. The bottom panel shows that the surface of the liquid is distorted by the magnet – the liquid oxygen is pulled up the side of the boiling tube.